\documentclass[twocolumn,aps,prl,showpacs,groupedaddress,floatfix,letterpaper,superscriptaddress,reprint]{revtex4}
\usepackage[utf8]{inputenc}
\usepackage{graphicx}
\usepackage{hyperref}
\usepackage{amsmath}
\usepackage{amsfonts}
\usepackage{siunitx} 
\usepackage{amssymb}
\usepackage{xspace}
\newcommand{\BePlus}{$^9$Be$^+$\xspace}

\begin{document}
\title{Robust and resource-efficient microwave near-field entangling $^9$Be$^+$ gate}

\author{G.~Zarantonello}
\affiliation{Institut für Quantenoptik, Leibniz Universität Hannover, Welfengarten 1, 30167 Hannover, Germany}
\affiliation{Physikalisch-Technische Bundesanstalt, Bundesallee 100, 38116 Braunschweig, Germany}
\author{H.~Hahn}
\affiliation{Institut für Quantenoptik, Leibniz Universität Hannover, Welfengarten 1, 30167 Hannover, Germany}
\affiliation{Physikalisch-Technische Bundesanstalt, Bundesallee 100, 38116 Braunschweig, Germany}
\author{J.~Morgner}
\affiliation{Institut für Quantenoptik, Leibniz Universität Hannover, Welfengarten 1, 30167 Hannover, Germany}
\affiliation{Physikalisch-Technische Bundesanstalt, Bundesallee 100, 38116 Braunschweig, Germany}
\author{M.~Schulte}
\affiliation{Institut für Theoretische Physik und Institut für Gravitationsphysik (Albert-Einstein-Institut), Leibniz Universität Hannover, Appelstrasse 2, 30167 Hannover, Germany}
\author{A.~Bautista-Salvador}
\affiliation{Institut für Quantenoptik, Leibniz Universität Hannover, Welfengarten 1, 30167 Hannover, Germany}
\affiliation{Physikalisch-Technische Bundesanstalt, Bundesallee 100, 38116 Braunschweig, Germany}
\affiliation{Laboratorium für Nano- und Quantenengineering, Leibniz Universität Hannover, Schneiderberg 39, 30167 Hannover, Germany}
\author{R.~F.~Werner}
\affiliation{Institut für Theoretische Physik, Leibniz Universität Hannover, Appelstrasse 2, 30167 Hannover, Germany}
\author{K.~Hammerer}
\affiliation{Institut für Theoretische Physik und Institut für Gravitationsphysik (Albert-Einstein-Institut), Leibniz Universität Hannover, Appelstrasse 2, 30167 Hannover, Germany}
\author{C.~Ospelkaus}
\affiliation{Institut für Quantenoptik, Leibniz Universität Hannover, Welfengarten 1, 30167 Hannover, Germany}
\affiliation{Physikalisch-Technische Bundesanstalt, Bundesallee 100, 38116 Braunschweig, Germany}
\affiliation{Laboratorium für Nano- und Quantenengineering, Leibniz Universität Hannover, Schneiderberg 39, 30167 Hannover, Germany}

\begin{abstract}
Microwave trapped-ion quantum logic gates avoid spontaneous emission as a fundamental source of decoherence. However, microwave two-qubit gates are still slower than laser-induced gates and hence more sensitive to fluctuations and noise of the motional mode frequency. We propose and implement amplitude-shaped gate drives to obtain resilience to such frequency changes without increasing the pulse energy per gate operation. We demonstrate the resilience by noise injection during a two-qubit entangling gate with $^9$Be$^+$ ion qubits. In absence of injected noise, amplitude modulation gives an operation infidelity in the $10^{-3}$ range.
\end{abstract}


\maketitle

Trapped ions are a leading platform for scalable quantum logic~\cite{brown_co-designing_2016,bermudez_assessing_2017} and quantum simulations~\cite{blatt_quantum_2012}. Major challenges towards larger-scale devices include the integration of tasks and components that have been so far only demonstrated individually, as well as single and multi-qubit gates with the highest possible fidelity to reduce the overhead in quantum error correction. Microwave control of trapped-ion qubits has the potential to address both challenges~\cite{mintert_ion-trap_2001,ospelkaus_trapped-ion_2008} as it allows the gate mechanism, potentially including control electronics, to be integrated into scalable trap arrays. Because spontaneous emission as a fundamental source of decoherence is absent and microwave fields are potentially easier to control than the laser beams that are usually employed, microwaves are a promising approach for high fidelity quantum operations. In fact, microwave two-qubit gate fidelities seem to improve more rapidly than laser-based gates. However, observed two-qubit gate speeds of laser-based gates~\cite{ballance_high-fidelity_2016,gaebler_high-fidelity_2016} are still about an order of magnitude faster than for microwave gates~\cite{hahn_integrated_2019-1,harty_high-fidelity_2016,weidt_trapped-ion_2016}. This makes gates more susceptible to uncontrolled motional mode frequency changes, as transient entanglement with the motional degrees of freedom is the key ingredient in multi-qubit gates for trapped ions. As other error sources have been addressed recently, this is of growing importance. Merely increasing Rabi frequencies may not be the most resource-efficient approach, as it will increase energy dissipation in the device. A more efficient use of available resources could be obtained using pulse shaping or modulation techniques. In fact, a number of recent advances in achieving high-fidelity operations or long qubit memory times have been proposed or obtained by tailored control fields. Examples include pulsed dynamic decoupling~\cite{manovitz_fast_2017}, Walsh modulation~\cite{hayes_coherent_2012}, additional dressing fields to increase coherence times~\cite{timoney_quantum_2011}, phase~\cite{milne_phase-modulated_2018}, amplitude~\cite{zhu_trapped_2006,roos_ion_2008,choi_optimal_2014,steane_pulsed_2014,palmero_fast_2017,schafer_fast_2018} and frequency modulation~\cite{leung_entangling_2018} as well multi-tone fields~\cite{haddadfarshi_high_2016,shapira_robust_2018-1,webb_resilient_2018}. In many cases, these techniques lead to significant advantages. For multi-qubit gates, one mechanism is to optimize the trajectory of the motional mode in phase space for minimal residual spin-motional entanglement in case of experimental imperfections. This effectively reduces the distance between the origin and the point in phase space at which the gate terminates in case of errors.

Here we propose and implement amplitude modulation for near-field microwave two-qubit entangling gates to make operations more resilient to normal mode frequency fluctuations, one of the dominant error sources in present experiments~\cite{hahn_integrated_2019-1}, without increasing the electrical energy cost per gate. We consider the bichromatic gate mechanism discussed in~\cite{molmer_multiparticle_1999,solano_deterministic_1999,milburn_ion_2000}. In a notation similar to~\cite{sorensen_entanglement_2000}, simultaneous application of blue and red motional sidebands of the qubit transition with detuning $\delta$ yields the propagator $U(t)= e^{-i A(t)S^2_y} e^{-i G(t)S_y x} e^{-i F(t)S_y p}$, where $x$ and $p$ are dimensionless position and momentum operators, $S_y=1/2(\sum_j \sigma^y_j)$ and $\sigma^y_j$ is the Pauli matrix for ion $j$. We have:
\begin{eqnarray}
\label{eq:1}
&F(t)=&-\sqrt{2}\int_0^t \Omega (t')\cos (\delta t')dt'\nonumber\\
&G(t)=&-\sqrt{2}\int_0^t \Omega (t')\sin (\delta t')dt'\\
&A(t)=&\sqrt{2}\int_0^t F(t')\Omega (t')\sin (\delta t')dt'\nonumber\,,
\end{eqnarray}
where $\Omega(t)$ is the time-dependent gate Rabi frequency. For eigenstates of $S_y$, $U(t)$ effectively leads to trajectories in phase space with dimensionless coordinates $G(t)$ and $F(t)$ for the harmonic oscillator of the motional mode. A closed trajectory is reached for $F(\tau)=G(\tau)=0$, where $\tau$ is the gate time. The final value of $A(\tau)$ is the area enclosed by the trajectory and thus the accumulated phase. As can be inferred from $U(t)$, the accumulated phase depends on the joint state of both ions and thus implements a two-qubit phase gate in the $S_y$ basis. In the $S_z$ basis, a maximally entangled state emerges from a product state for $|A(\tau)|=\pi/2$. We introduce the dimensionless envelope $P(t)$ through $\Omega(t)=\Omega_\mathrm{MS} P(t)$. For $P(t)$ constant in the range $0\le t\le\tau$ and $0$ otherwise, one obtains the well-known square pulse gate.

Consider the class of functions $P(t)=\sin^n(\alpha t)$ with $\alpha$ and $n$ suitable constants, which also ensure $P(t)=0$ at the beginning and a `soft' start~\cite{sutherland_versatile_2019}. For near-field microwave gates, `soft' start in amplitude modulation is known to suppress unwanted motional excitation from microwave electric pseudopotential kicks~\cite{warring_techniques_2013}. At the total gate time $\tau$, a `soft' end is desirable, which implies $\alpha \tau=m\pi$ for an integer $m$ identifying the number of pulses present in the envelope. At the end the phase space loop also needs to be closed, which puts a constraint on $\delta$. Without losing generality, we restrict ourselves to the case of $n=2$ and $m=1$. The integrals~(\ref{eq:1}) can be solved analytically, and one finds that multiple sets of $\tau$ and $\delta$ yield the required gate phase $|A(\tau)|=\pi/2$ and a closed trajectory. 
\begin{figure}[tb]
	\centering
	\includegraphics[width=1.0\columnwidth]{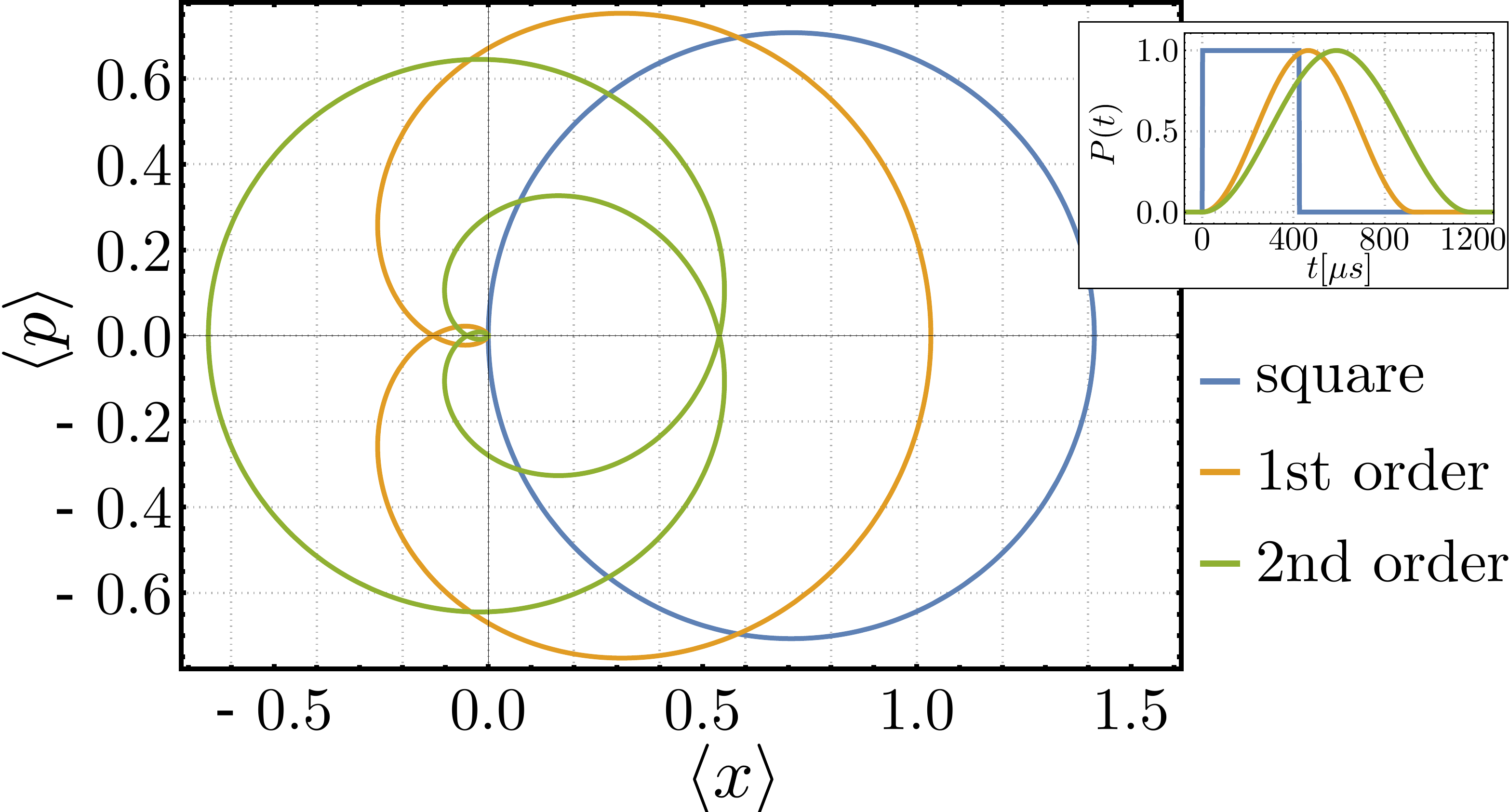}
	\caption{
	Phase space trajectories for a representative spin state in case of a square pulse gate (blue), a first ($k=1$; orange) and second ($k=2$; green) order amplitude modulated gate using a $\sin^2$ amplitude modulation function. The inset shows the three envelopes which produce the main plot trajectories for $\Omega_\mathrm{MS}/2 \pi = 1.18\,$kHz. 
	}
	\label{fig:figure1}
\end{figure}	
For $n=2$ and $m=1$ the detuning is:
\begin{equation}
\label{eq:2}
\delta_k=\frac{2\pi(k+1)}{\tau_k}\nonumber\,,
\end{equation}
where $k$ is the order of the shaped gate and $\tau_k$ the gate time required to generate the maximally entangled state for this order. The latter can be calculated analytically using equations~(\ref{eq:1}) and the constraints mentioned above. Figure~\ref{fig:figure1} shows the phase space trajectories of a representative spin state for the square pulse gate and for the first two orders of $\sin^2$ amplitude modulation. Increasing orders will exhibit more windings with a reduced radius around the origin. In general, this reduced radius will alleviate the impact of symmetric errors such as a miscalibrated secular mode frequency $\omega_r$ or detuning $\delta$. This is because $F(\tau)$ and $G(\tau)$, in the presence of errors, end up closer to the phase space origin than for the square pulse, therefore more reliably disentangling the qubit degree of freedom from the motional state~\cite{haddadfarshi_high_2016}.

We use \BePlus ions in a surface-electrode trap with integrated microwave conductors described in~\cite{hahn_integrated_2019-1}. Doppler cooling and detection are performed on the closed-cycle transition $\mathrm{^{2}S_{1/2}}\left|F=2,m_F=2\right>$ $\leftrightarrow$ $\mathrm{^{2}P_{3/2}}\left|m_J=\frac{3}{2},m_I=\frac{3}{2}\right>$ at $\lambda=313\,$nm; the detection window is $400\,$\SI{}{\micro\second} long. We use the hyperfine transition in the electronic ground state $^{2}S_{1/2}$ $\left|F=2,\,m_F=+1\right>\equiv\left|\uparrow\right>$ $\leftrightarrow$ $\mathrm{^{2}S_{1/2}}\left|F=1,\,m_F=+1\right>\equiv\left|\downarrow\right>$ as our qubit, which for a magnetic field of $\left|\mathrm{\bf{B_0}}\right| \simeq 22.3 \,\mathrm{mT}$ has a frequency of $\omega_0\simeq 2 \pi \times 1082.55\,\mathrm{MHz}$ and is first-order field-independent, allowing long coherence times~\cite{langer_long-lived_2005}. $F$ is the total angular momentum, $J$ the total electronic angular momentum, $I$ the nuclear spin, $m_F$, and $m_J$ and $m_I$ their respective projections on the quantization axis. All carrier transitions in the $^{2}S_{1/2}$ manifold are excited by resonant microwaves from a conductor embedded in the trap. To perform high-fidelity carrier operations we use composite pulses sequences~\cite{torosov_smooth_2011,genov_correction_2014} to realize $\pi$ and $\pi/2$ rotations for state preparation, shelving and analysis. 

Sideband transitions are excited using a single microwave conductor designed to produce a strong oscillating magnetic field quadrupole~\cite{wahnschaffe_single-ion_2017} at the desired frequency. The quadrupole is designed to provide the gradient necessary for spin-motion coupling while reducing the residual field at its minimum to avoid off-resonant carrier excitation. By applying a microwave power of $\sim5.5\,\mathrm{W}$, we obtain a gradient of around $19\,\mathrm{T/m}$. Microwave amplitude modulation is performed by an arbitrary waveform generator~\cite{bowler_arbitrary_2013} providing the setpoint of a digital PI controller~\cite{hannig_highly_2018} which in turn controls a fast analog multiplier~\cite{hahn_two-qubit_2019}.

For the bichromatic microwave gate drive, we measure $\Omega_\mathrm{MS}/2\pi=1.18\,$kHz. The gate is carried out on the two-ion low frequency (LF) out-of-phase radial mode at a frequency of $\omega_r/2\pi=6.16\,$MHz. Using sideband thermometry~\cite{turchette_heating_2000}, we estimate an average occupation of $\bar{n}=0.4(1)$ and a heating rate of $\dot{\bar{n}} = 8.4(7)\,\mathrm{s}^{-1}$. Throughout this work, no `warm-up' pulse was employed to compensate the effect of an observed `chirp' in the motional mode frequency~\cite{harty_high-fidelity_2016,hahn_integrated_2019-1}, likely caused by thermal transients from microwave currents in the trap. This avoids additional energy dissipation not strictly related to gate operation.

Amplitude modulation of the driving fields affects not only the gate Rabi rate but potentially also the qubit energy splitting through power-dependent shifts, such as the differential AC Zeeman shift. This shift arises from non-zero oscillatory fields that accompany the oscillating gradient and introduce a new time-dependent term in the Hamiltonian
\begin{equation}
H_{Z}(t)= P^2(t)\frac{\Delta}{2} \sum_j \sigma^z_j\,,
\end{equation}
where $\Delta$ is the peak differential AC Zeeman shift, and $\sigma^z_j$ is the Pauli matrix for ion $j$. Experimentally, this can be addressed in several ways, one of which is to drive the gate using sidebands tones with time-dependent frequencies $\omega(t)=\omega_0+\Delta P^2(t)\pm(\omega_r+\delta)$, where the sign identifies the blue or red sideband. Another possibility is dynamic decoupling~\cite{harty_high-fidelity_2016}. Here we employ a microwave conductor designed to minimize the residual field at the ion position and hence make $\Delta$ as small as possible.

The AC Zeeman shift induced by a single sideband on our qubit transition is dominated by the projection of the microwave field on $\mathbf{B_0}$ ($\pi$ component). For the bichromatic drive, the shifts due to the $\pi$ components of the two sidebands would have opposite sign and ideally cancel each other. Any remaining shift is due to off-resonant coupling to $\Delta m_F=\pm 1$ transitions detuned by $\approx 200\,\mathrm{MHz}$ from the qubit and induced by the microwave's field projection orthogonal to $\mathbf{B_0}$ ($\sigma$ components). The trap is engineered to have a minimum of the oscillating magnetic field as close to the pseudopotential null as possible. Due to imperfections, it is displaced from the pseudopotential null by about~$1.5\,$\SI{}{\micro\meter}. We operate our gate close to this position, where the observed AC Zeeman shift on the qubit transition is minimized (the $\sigma$ field components effectively vanish, giving $\Delta \le 5\,$Hz). Because of the increased micromotion, the ions are driven away from the minimum periodically at a rate given by the RF drive frequency. Because of the spatial dependence of the AC Zeeman shift around the chosen position, an additional time-dependent shift may then occur.

\begin{figure}[tb]
	\centering
	\includegraphics[width=0.85\columnwidth]{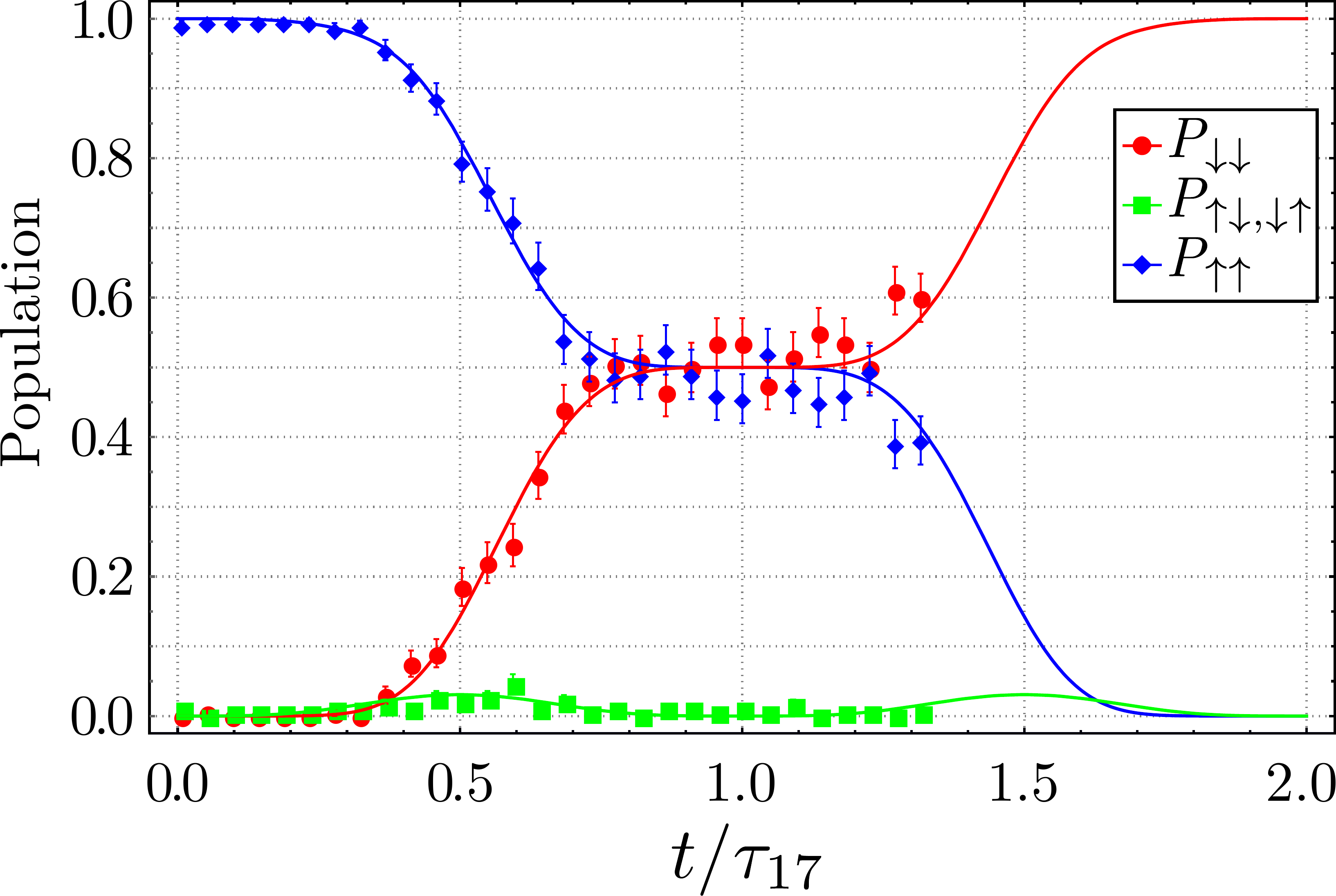}
	\caption{
		Dynamics of the internal state of the ions during the time evolution of the $\sin^2$ pulse. The flat region around $t/\tau_{\mathrm{17}}=1$ is expected by the analytical model (solid lines) and observed in the experiment. Each data point is an average of 200 experiments. 
	}
	\label{fig:figure2}
\end{figure}

In general the Bell state fidelity $\mathcal{F}$ is a function of $F$, $G$ and $A$~\cite{shapira_robust_2018-1}. For the $\sin^2$ pulse, we find that all derivatives of the fidelity $\frac{\partial^{n} \mathcal{F}(F,G,A)}{\partial t^{n}}|_{t=\tau_k}$ in the motional ground state, $n=0$, are equal to $0$, demonstrating the intrinsic resilience against timing imperfections. One can observe this behavior by turning off the microwave drive at different times of the $\sin^2$ pulse. Figure~\ref{fig:figure2} shows experimental data for $k=17$ $(\tau_{\mathrm{17}}=2938\,\SI{}{\micro\second})$ together with predictions from the analytic solution of~\cite{sorensen_entanglement_2000}. As expected from the derivatives of $F(t)$, $G(t)$ and $A(t)$, the population dynamics is stable around $t=\tau_{\mathrm{17}}$, where the derivatives vanish.

To compare the performance of the amplitude modulated gate to a square pulse gate, a relevant quantity for microwave near-fields is given by the total energy deposited in the trap structure by the bichromatic current, due to potential thermal effects. This is different from laser-based gates, where available laser power typically imposes limits to gate speeds. We therefore compare the gate fidelity to a square pulse gate with seven loops in phase space and $\tau=1122\,$\SI{}{\micro\second} since the pulse energies are equal. From finite element simulations~\cite{hahn_multilayer_2019}, the microwave conductor reflects $91.1\,$\% of the amplitude; the energy dissipated per gate is about $1\,$mJ.
To prove the resilience in a direct comparison, we amplitude modulate the RF trap drive with Gaussian noise~\cite{Noise_source}, thereby introducing fluctuations of the radial mode frequency. To characterize the amount of noise injected, we measure the instantaneous linewidth of $\omega_r$ for different values of the noise source's amplitude: after resolved sideband cooling to near the motional ground state, we excite the motion with a weak near-resonant electric field, and apply a red sideband $\pi$ pulse to flip the spin conditional on the motional excitation. The FWHM of the signal as a function of the electric field frequency is taken as a measure of the injected noise. Gates are carried out in an interleaved way between the $\sin^2$ and the square pulse gate, in order to probe the same conditions for both amplitude shapes. The fidelity of the maximally entangled state, $1/\sqrt{2}(\left|\uparrow\uparrow\right>-i\left|\downarrow\downarrow\right>)$, is extracted from parity oscillations and from the fluorescence signal of the $P_{\uparrow\uparrow}$ and $P_{\downarrow\downarrow}$ signal generated by scanning the phase of a $\pi$/2 analysis pulse~\cite{sackett_experimental_2000}. Here we determined the state populations using a sum of weighted Poissonians as in~\cite{hahn_integrated_2019-1}. 

\begin{figure}[tb]
	\centering
	\includegraphics[width=1.\columnwidth]{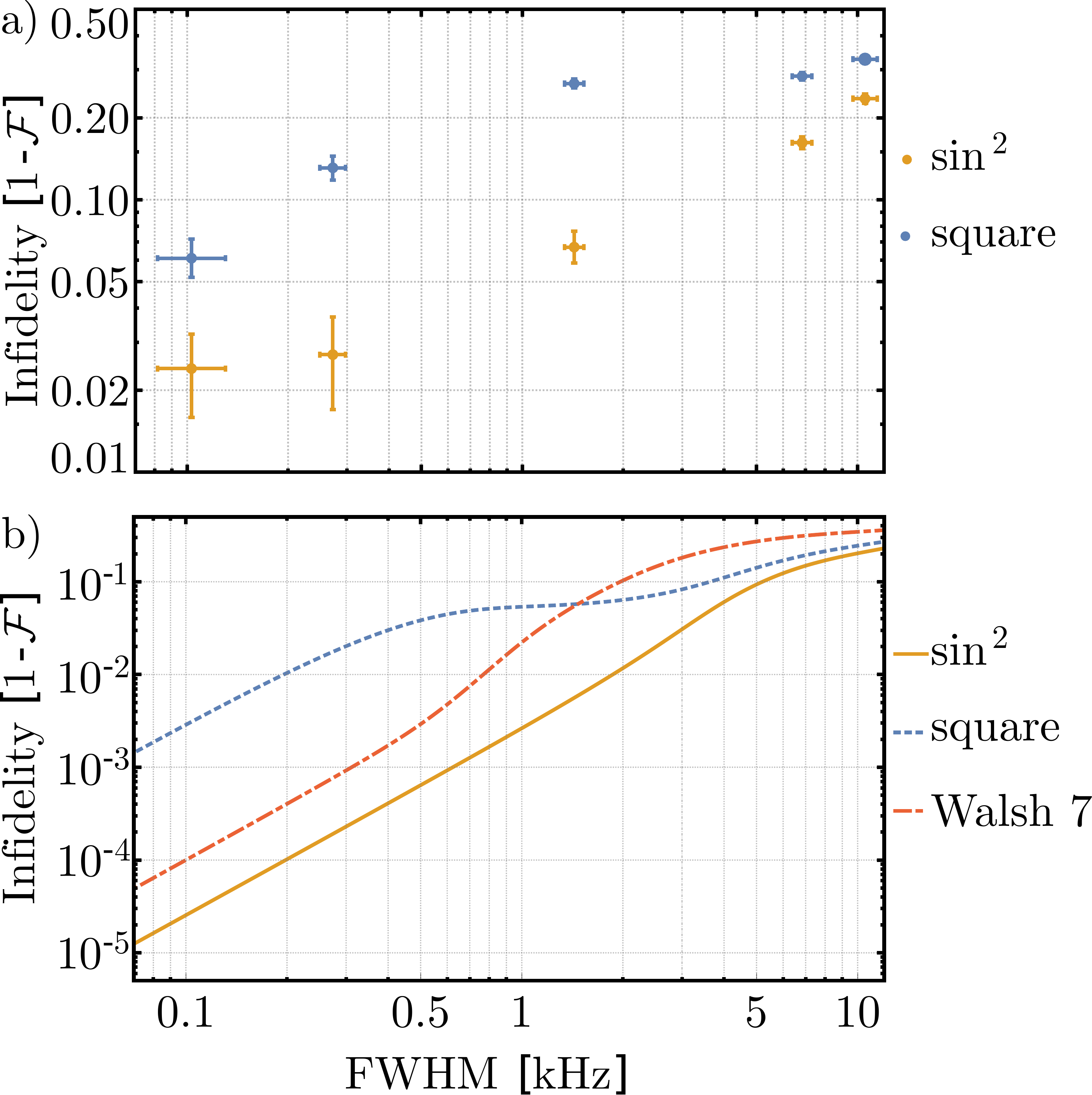}
	\caption{
		Comparison between square pulse and $\sin^2$ amplitude modulated gate: a) Experimental result; at any level of injected noise, the amplitude modulated gate with $k=17$ results in higher fidelities than the $7$ loop square pulse gate; b) Expected infidelity from an analytic model for different schemes: $\sin^2$ with $k=20$, $8$ loop square pulse, Walsh[7,x] modulation on 8 loop gate. The model accounts only for initial $\bar{n}=0.4$; no other error sources have been included.
	}
	\label{fig:figure3}
\end{figure}

The measured fidelities as a function of the radial mode FWHM due to the injected noise are shown in Fig.~\ref{fig:figure3}a) and indicate that the amplitude modulated scheme suffers considerably less from noise than the standard square pulse scheme. In the latter case, reported fidelities are lower than theoretically expected, mainly due to slow drifts of the mode during data acquisition (about six minutes for data shown). The effect of slow variations of $\omega_r$ during the different acquisitions of the scan is different for each datapoint due to varying experimental conditions and therefore cannot be replicated accurately by theory. Fig.~\ref{fig:figure3}b) shows the expected infidelity for different schemes using an analytic model. We compare a standard 8 loop square pulse scheme which requires $\tau=1200\,$\SI{}{\micro\second} with a $k=20$ $\sin^2$ modulation, $\tau_{\mathrm{20}}=3200\,$\SI{}{\micro\second}. The two schemes have been chosen since they have the same pulse energy. We also compare to an improved version of the standard scheme which makes use of Walsh[7,x] modulation~\cite{hayes_coherent_2011} on the 8 loop gate. The amplitude modulated scheme at the same pulse energy presents a lower infidelity.

\begin{figure}[tb]
\centering
\includegraphics[width=1.0\columnwidth]{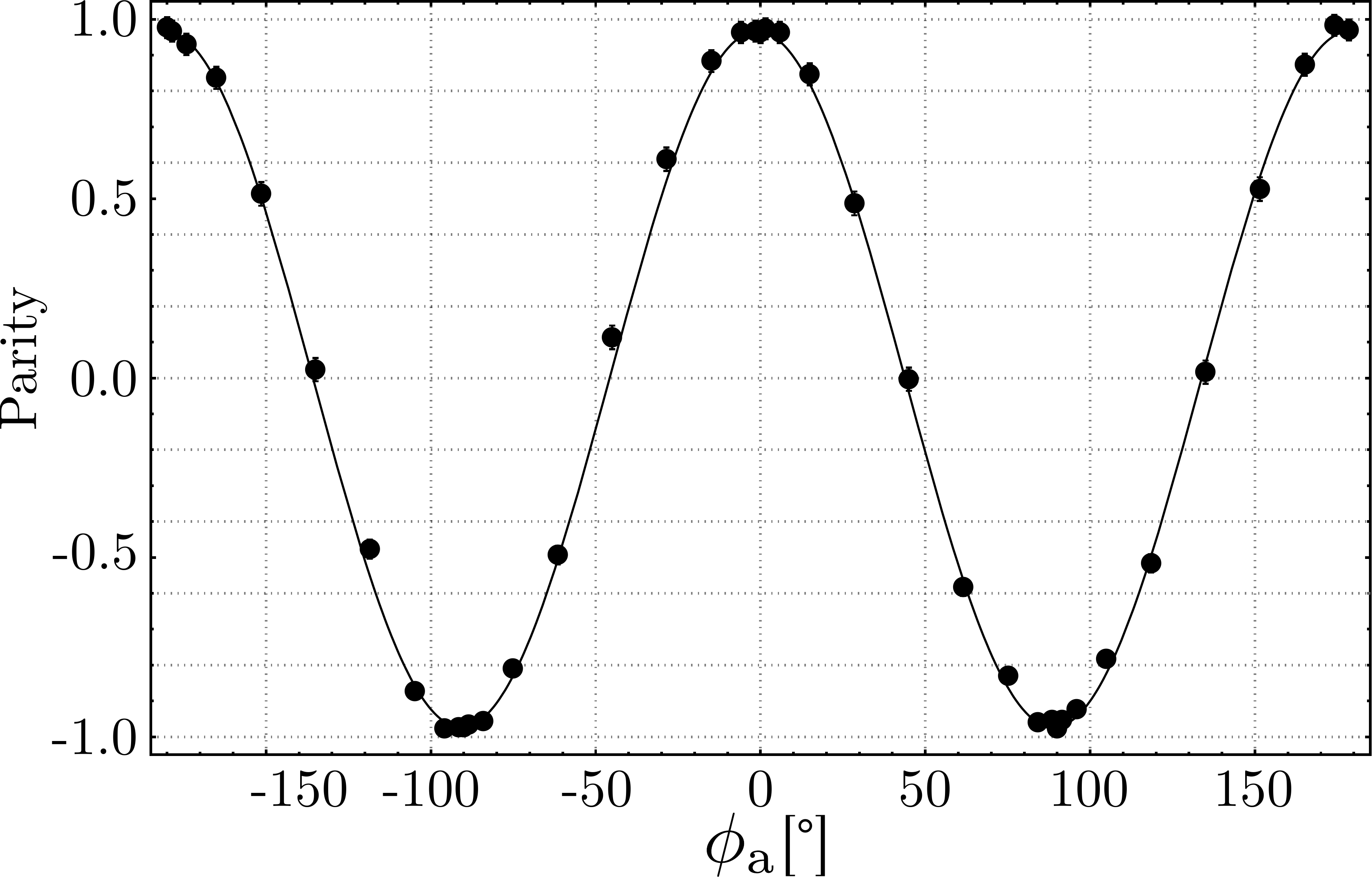}
\caption{
	Parity oscillations for $\sin^2$ shaped gate obtained by determining the state populations with thresholds in the fluorescence histogram. Each point is an average of 600 experiments. 
}
\label{fig:figure4}
\end{figure}

The best gate fidelities are obtained without the noise injection system attached, using a $\sin^2$ $k=17$ pulse, under the same experimental conditions stated previously, and with a circuit to improve the stability of the delivered RF power (similar to~\cite{harty_high-fidelity_2013}). The dataset is composed of two consecutive scans of the phase of the analysis $\pi$/2 pulse to extract parity oscillations. Here each phase is probed $300$ times. To obtain reference histograms we detect and prepare four states, each measured $2\times 10^4$ times. 
The Bell state fidelity is estimated using three methods. For the first method, state populations are determined using the sum of weighted Poissonians mentioned previously. To estimate the fidelity we perform a resampling bootstrap analysis. We generate multiple synthetic datasets by randomly assembling the data in two separate scans where from one we extract the populations $P_{\uparrow\uparrow}$, $P_{\downarrow\downarrow}$ and from the other the parity amplitude. The operation is repeated $1000$ times, resulting in a distribution of fidelities. We obtain a fidelity $\mathcal{F}=99.5$\% with a $68$\% confidence interval of $[99.3 , 99.7]$\%. For the second method, the populations are determined by dividing the fluorescence histograms using appropriate thresholds into three bins (i.e.\ zero, one or two ions bright~\cite{harty_high-fidelity_2016}). The resulting bootstrapped fidelity distribution has mean $\mathcal{F}=99.7$\% (SPAM error corrected with $\epsilon_\mathrm{SPAM}=1.5(1)\,$\%) with $68$\% confidence interval $[99.6 , 99.8]$\%. Fig.~\ref{fig:figure4} shows the combined parity oscillations from the original sets of data derived from the threshold analysis. Finally, the third method to extract the fidelity is the maximum-likelihood algorithm described in~\cite{keith_joint_2018}. With a training fraction of $20\%$ and a bootstrap of $1000$, a fidelity of $\mathcal{F}=99.2\,$\% with a bootstrapped $68$\% confidence interval $[99.1 , 99.7]$\% is inferred. The uncertainty is larger because this algorithm produces a joint uncertainty on state analysis and tomography, whereas the two former methods estimate the fidelities after the states have already been assigned to the raw data. In the limit of vanishing SPAM error, the two former and the latter method should yield comparable uncertainties.

We now expect a major contribution to the error budget to be imperfections in the assumption $\Delta \simeq 0$. On one hand, time-varying shifts in the ion position relative to the AC Zeeman shift minimum, induced by fluctuating stray potentials, may cause variations of the AC Zeeman shift. On the other hand, as previously mentioned, micromotion can also lead to additional time-dependent AC Zeeman shifts. The strongest variation of the differential AC Zeeman shift expected from our finite element simulations is $0.6\,$Hz/nm. Assuming that one ion exhibits an AC Zeeman shift of $20\,$Hz relative to the other, which is at $0\,$Hz shift, simulations predict an infidelity of $1.1\times10^{-3}$. We expect gate infidelity contributions from motional heating of $\approx2\times10^{-4}$, from imperfect ground state cooling of $\approx1\times10^{-5}$ and of $<1\times10^{-5}$ from the motional frequency `chirp'. Spectator modes contribute a simulated error of $5\times10^{-4}$, which can be mitigated by better engineering of trap potentials or by exploring additional modulation schemes designed to address spectral crowding~\cite{leung_robust_2018}.

In summary, we have introduced amplitude modulated two-qubit microwave near-field gates and demonstrated their resilience to motional mode changes compared to the standard square pulse gate with the same pulse energy using noise injection, thereby addressing the major current challenge for these types of gates. The fully optimized gate reaches an infidelity in the $10^{-3}$ range. It might be useful to evaluate other pulse shapes such as Blackman pulses, weighted series of sines with different $\alpha$ or even piecewise functions with a sufficient number of steps as already implemented with lasers~\cite{figgatt_parallel_2019}. Solutions to the remaining AC Zeeman shifts comprise better engineering of the magnetic field quadrupole, aimed at minimizing the differential AC Zeeman shift rather than the residual magnetic field at the minimum. The technique presented here is compatible with continuous dynamical decoupling~\cite{harty_high-fidelity_2016} which would also allow to reduce this source of error. An interesting perspective to further increase the gate speed would be the combination with motional squeezing~\cite{ge_trapped_2019}.

\begin{acknowledgments}
We thank P.~O.~Schmidt and S.~A.~King for helpful discussions and comments on the manuscript. We are grateful for discussions and suggestions from T.~P.~Harty, R.~Jördens, D.~H.~Slichter and R.~Srinivas.
We acknowledge funding from the clusters of excellence `QUEST' and `Quantum Frontiers', from the EU QT flagship project `MicroQC' and from DFG through CRC 1227 DQ-\textit{mat}, projects A01, A05 and A06 and from PTB and LUH.
\end{acknowledgments} 

\bibliographystyle{apsrev4-1}
\bibliography{qc}

\end{document}